\documentstyle[prl,aps,psfig]{revtex}
\newcommand{\beq}{\begin{equation}}
\newcommand{\eeq}{\end{equation}}
\newcommand{\beqa}{\begin{eqnarray}}
\newcommand{\eeqa}{\end{eqnarray}}
\newcommand{\ba}{\begin{array}}
\newcommand{\ea}{\end{array}} 

\begin{document} 
\draft 

%%%%%%%%%%%%%%%%%%
\twocolumn[\hsize\textwidth\columnwidth\hsize\csname
@twocolumnfalse\endcsname
%%%%%%%%%%%%%%%%%%

\widetext 

\title{Thermodynamics of Solitonic Matter Waves in a Toroidal Trap} 
\author{L. Salasnich$^{1}$, A. Parola$^{2}$, and L. Reatto$^{3}$} 
\address{
$^{1}$CNR-INFM and CNISM, Unit\`a di Milano Universit\`a, 
Via Celoria 16, 20133 Milano, Italy \\ 
$^{2}$Dipartimento di Fisica e Matematica, 
Universit\`a dell'Insubria, \\ 
Via Valleggio 11, 22100 Como, Italy\\
$^{3}$Dipartimento di Fisica and CNISM, Universit\`a di Milano, \\ 
Via Celoria 16, 20133 Milano, Italy} 

\maketitle 

\begin{abstract} 
We investigate the thermodynamic properties of
a Bose-Einstein condensate with negative scattering
length confined in a toroidal trapping potential.
By numerically solving the coupled Gross-Pitaevskii and
Bogoliubov-de Gennes equations,
we study the phase transition from the uniform state
to the symmetry-breaking state characterized by a
bright-soliton condensate and a localized thermal cloud.
In the localized regime three states with a finite condensate
fraction are present: the thermodynamically stable
localized state, a metastable localized state and also
a metastable uniform state. Remarkably, the presence of
the stable localized state strongly increases the critical temperature
of Bose-Einstein condensation.
\end{abstract} 


%%%%%%%%%%%%%%%%%%
]
%%%%%%%%%%%%%%%%%%

\narrowtext 

In recent experiments a repulsive Bose-Einstein condensate (BEC) 
has been produced and studied in a quasi one-dimensional (1D) 
ring \cite{stamper05,arnold}. 
These experiments and also previous experimental investigations 
with cold atoms in torodial traps \cite{otherexp} are important steps 
to achieve ultra-high precision sensors with atom interferometry 
(see for instance \cite{kasevich}). 
The case of an attractive BEC in a ring has not yet been 
experimentally investigated but appears very interesting:  
a quantum phase transition from an azimuthally uniform condensate to a bright soliton 
has been predicted \cite{jap1,jap2}. 
%Moreover, we have theoretically shown that the system supports also 
%which are energetically unstable but can be 
%dynamically stable \cite{sala05}. 
Dynamically stable multi-peak bright solitons may also appear \cite{sala05}.
\par 
In this paper we investigate the effect of 
temperature on the uniform-to-localized phase transition 
by solving the Gross-Pitaevskii equation (GPE) 
of the macroscopic 
wave function of the Bose condensate and the Bogoliubov-de Gennes 
equations for the quantum depletion and the thermal cloud. 
We show that, for a fixed number $N$ of atoms, 
the uniform solution always exists 
but it is thermodynamically stable only up to a critical 
interaction strength which depends on the temperature. 
Above this critical interaction strength both condensate and 
thermal cloud become localized. In addition, 
we determine the finite-temperature 
phase-diagram of the system for a sample of alkali-metal atoms 
in a ring. Our results are relevant not only for the 
physics of cold atoms but also for the 
nonlinear science of solitons. In fact, a mixture of condensed 
and non condensed attractive particles \cite{vardi} 
correspond to incoherent solitons of light in Kerr media 
\cite{mitchell}. 
\par 
We consider a Bose gas with negative scattering 
length ($a_s<0$) confined in a toroidal potential 
and model the transverse confinement with a harmonic 
potential of frequency $\omega_{\bot}$. 
The two characteristic lengths of the toroidal 
trap are the azimuthal radius $R$ and the 
transverse harmonic length $a_{\bot}=(\hbar/(m\omega_{\bot}))^{1/2}$. 
To avoid the confinement-induced resonance 
at $|a_s|\simeq a_{\bot}$ \cite{olshanii} 
we impose that $|a_s| \ll a_{\bot}$ \cite{sala}. 
At low temperature $T$ and with a finite number $N$ of atoms, 
in the Bose gas there is the thermal cloud but also the BEC. 
Under the condition $R\gg a_{\bot}$ 
the azimuthal wavefunction $\phi(z)$ of the BEC in the ring 
satisfies 
the 1D non-polynomial Schr\"odinger equation (NPSE) 
\cite{sala05,npse} 
\beq 
\left[ 
-{\hbar^2\over 2m}\partial_z^2 + \mu(n_0(z)) \right] 
\phi(z) = {\bar \mu} \; \phi(z) \; , 
\eeq 
where $z=R\theta$ is the azimuthal coordinate with $\theta$ 
the azimuthal angle. This equation has been deduced 
from the 3D GPE by using a Gaussian transverse 
wave function with a density-dependent width 
and neglecting curvature effects \cite{sala05,npse}. 
The non-polynomial term $\mu(n_0(z))$ 
is a function of the BEC density $n_0(z)=N_0|\phi(z)|^2$ 
and it is given by 
$
\mu = { \partial{\cal E}/ \partial n_0 } 
$,
where 
$
{\cal E}= \hbar \omega_{\bot} \; n_0(1 + 2 a_s n_0)^{1/2} 
$
and $a_s<0$ is the 3D s-wave scattering length \cite{sala05}. 
The wave function $\phi(z)$ is normalized to one and satisfies 
the condition of periodicity $\phi(z+L) = \phi(z)$, 
where $L=2\pi R$. The chemical potential ${\bar \mu}$ is fixed 
by the normalization condition 
$
\int |\phi(z)|^2 dz = 1  
$. 
Quantum and thermal depletions of the BEC 
are obtained by solving 
the Bogoliubov-de Gennes (BdG) equations \cite{fetter} 
for the quasi-particle amplitudes $u_{\alpha}(z)$ 
and $v_{\alpha}(z)$, given by 
\beq 
{\cal L}
\left( 
\ba{c} 
u_{\alpha}(z) \\
v_{\alpha}(z) 
\ea
\right) 
= \epsilon_{\alpha} 
\left( 
\ba{c} 
u_{\alpha}(z) \\
v_{\alpha}(z)  
\ea
\right) \; , 
\eeq
where $\epsilon_{\alpha}$ are the energies of quasi-particle 
excitations and the operator ${\cal L}$ is 
\beq 
{\cal L} = 
\left( 
\ba{cc} 
-{\hbar^2\over 2m}\partial_z^2 - {\bar \mu} + 
{\partial (n_0 \mu ) \over \partial n_0} 
& 
{N_0\phi^2} {\partial \mu \over \partial n_0} 
\\
- {N_0(\phi^*)^2} {\partial \mu \over \partial n_0} 
& 
{\hbar^2\over 2m}\partial_z^2 +{\bar \mu} - 
{\partial (n_0 \mu ) \over \partial n_0} 
\ea
\right) \; .   
\eeq 
The local density of the quantum depletion reads 
$n_{out}(z) = \sum_{\alpha} |v_{\alpha}(z)|^2$,  
while the thermal local density is 
$n_T(z) = \sum_{\alpha} \left( |u_{\alpha}(z)|^2 
+ |v_{\alpha}(z)|^2 \right) \bar{n}_{\alpha}$, 
where 
$\bar{n}_{\alpha} = 
\left( \exp(\epsilon_{\alpha}/k_B T) - 1 \right)^{-1}$ 
are the Bose numbers of occupation for the non-interacting 
quasi-particles, $k_B$ is the Boltzmann constant and $T$ is the 
absolute temperature of the thermal cloud. 
The number ${\tilde N}$ of non-condensed atoms is thus 
given by ${\tilde N} = N_{out} + N_T$, where 
$N_{out} = \int n_{out}(z) dz$ and $N_T = \int n_T(z) dz$. 
In the weak-coupling limit, where $a_s n_0\ll 1$, the energy 
density ${\cal E}$ becomes ${\cal E}=\hbar\omega_{\bot}n_0 
+ \hbar\omega_{\bot}a_s n_0^2$, 
the NPSE (1) reduces to the familiar 1D GPE \cite{sala05} 
and the BdG equations (2,3) reduce to the BdG equations 
of the strictly 1D problem. 
Note that the Eqs. (1) and (2,3) could be improved by 
taking into account Popov and Beliaev corrections, 
which are usually not negligible only when the condensed 
fraction is very small \cite{griffin}.  
\par 
In our approach, by fixing the 
total number $N = N_0 + {\tilde N}$ of atoms 
and the temperature $T$, 
the equilibrium configurations are found 
from the Helmholtz free energy \cite{fetter} 
\beq 
F = E_0 + {\tilde E} + {\bar \mu} (N - N_0 ) - T S \; , 
\eeq 
where 
\beq 
E_0 = N_0 \int 
\left[ -{\hbar^2\over 2m}
\phi^*(z)\partial_z^2 \phi(z) + {\cal E}(n_0(z)) \right]  dz 
\eeq
is the BEC energy. The out-of-condensate 
energy $\tilde E$ and entropy $S$ are instead given by 
\begin{eqnarray} 
{\tilde E} &=& \sum_{\alpha} \epsilon_{\alpha} 
\left[ \bar{n}_{\alpha} - \int |v_{\alpha}(z)|^2 dz \right]  \\
S &=& -k_B \sum_{\alpha} \left[ 
\bar{n}_{\alpha} \log( \bar{n}_{\alpha} ) - 
(1 + \bar{n}_{\alpha} ) \log( 1 + \bar{n}_{\alpha} ) 
\right] \; . 
\end{eqnarray}
The solution of the 1D GPE with periodic boundary conditions 
is known analytically \cite{jap1}: it is uniform for 
$0<\gamma N_0 <\pi^2 a_{\bot}/L$, with $\gamma = 2 |a_s|/a_{\bot}$ 
the interatomic strength, and becomes localized 
for $\gamma N_0 > \pi^2 a_{\bot}/L$. By using the NPSE we find 
that the transition strength approaches the 1D GPE one 
for large $L$ \cite{sala05}. 
The main difference between 1D GPE and NPSE is that the NPSE 
correctly gives the collapse of the localized solution for 
$\gamma N_0 > 4/3$, while the 1D GPE does not predict 
collapse \cite{sala05}. It is important to observe that 
the uniform-to-localized transition depends on 
$N_0$ and only implicitly on $N=N_0+N_{out}+N_T$. 
We shall show that, for a fixed number $N$ of atoms, 
the uniform solution exists for any value of the interatomic 
strength $\gamma$ but it is thermodynamically 
stable only below a critical strength $\gamma_c$. Above 
$\gamma_c$ the stable state is a localized solution 
that minimizes the free energy. 
\par 
The BdG equations (2,3) are easily solved 
if $\phi(z)$ is uniform: $\phi(z)=1/\sqrt{L}$. In this case 
the quasi-particle amplitudes $u_k(z)=\bar{u}_k e^{ikz}$ 
and $v_k(z)=\bar{v}_k e^{-ikz}$ are 
plane waves and the wave vector $k$ is quantized: $k=(2\pi/L)j$, 
with $j$ an integer number. 
When $\phi(z)$ is not uniform, one must numerically solve 
Eq. (1) and Eqs. (2,3). We use a finite-difference 
space discretization and diagonalize the matrix associated 
to the operator $\cal L$ of Eq. (3). In our calculations 
we use various matrix dimensions up to $4000\times 4000$. 
The numerical procedure is discussed in \cite{sala99}. 
In our approach the transverse excitations 
do not contribute to the thermodynamics: this assumption 
is reasonable only if the system is quasi-1D. 
\begin{figure}
\centerline{\psfig{file=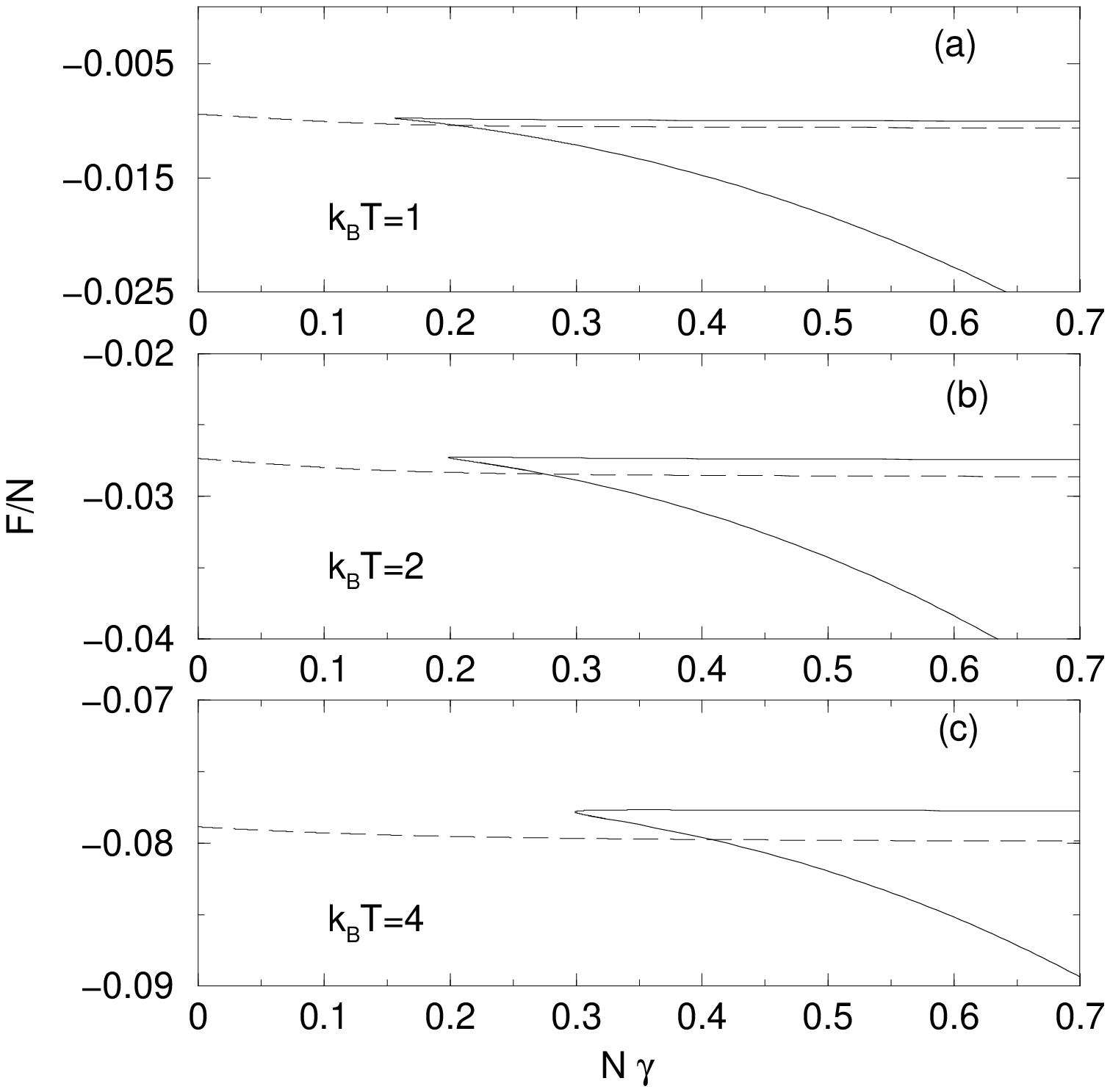,height=2.9in}}
{FIG. 1 Free energy $F$ as a function 
of the scaled interatomic strength $\gamma = 2|a_s|/a_{\bot}$ 
for an attractive Bose gas of $N=10^4$ atoms 
in a toroidal trap with $L/a_{\bot}=100$. 
Dashed line: uniform solution. 
Solid line: localized solutions. 
The energies $F$ and $k_BT$ are 
in units of $\hbar \omega_{\bot}$. } 
\end{figure}

\par 
In Fig. 1 we plot the results of our numerical calculations 
of the free energy $F$ as a function of the scaled 
interatomic strength $\gamma=2|a_s|/a_{\bot}$ 
for increasing values of the temperature $T$. 
We choose $N=10^4$ atoms and a toroidal geometry 
with $L/a_{\bot}=100$. While the uniform 
solution (dashed line) exists for any strength $\gamma$,  
the localized solution (solid line) exists only above a 
critical strength $\gamma_e$. As expected, 
at $T\simeq 0$ we find $\gamma_c \simeq 
\pi^2a_{\bot}/L=0.0987$. For a given $\gamma$ ($\gamma >\gamma_e$) 
there are two localized states: one is thermodynamically 
stable (lower solid line) and the other is metastable 
(upper solid line). The metastability is guaranteed 
by the Bogoliubov eigenvalues being real. 
At very-low temperature, see panel (a) of Fig. 1, 
for $\gamma > \gamma_e$ the solid line 
of the metastable localized solution is very close to the 
dashed line of the metastable uniform solution. 
At higher temperature, see panels (b) and (c) of Fig. 1, 
the free energy of the metastable localized solution, but also
the free energy of the stable one at small $\gamma > \gamma_e$, 
is higher than the free energy of the uniform solution. 
The strength $\gamma_c$ at which 
the dashed line crosses the lower solid line determines the 
point where the first-order transition takes place. 
At $T\simeq 0$ we find that $\gamma_c \simeq \gamma_e$. 
The strength $\gamma_c$ grows with the temperature $T$ 
but in general $\gamma_c$ does not coincide 
with the strength $\gamma_e$ at which the 
localized solutions appear. Thus, at finite temperature, 
for $0 < \gamma <\gamma_e$ there is only the stable uniform state; 
for $\gamma_e < \gamma < \gamma_c$ there are two metastable 
localized states and the stable uniform state; 
for $\gamma > \gamma_c$ the stable state is one of the 
two localized state while the other localized state and the 
uniform state are metastable. 
\begin{figure}
\centerline{\psfig{file=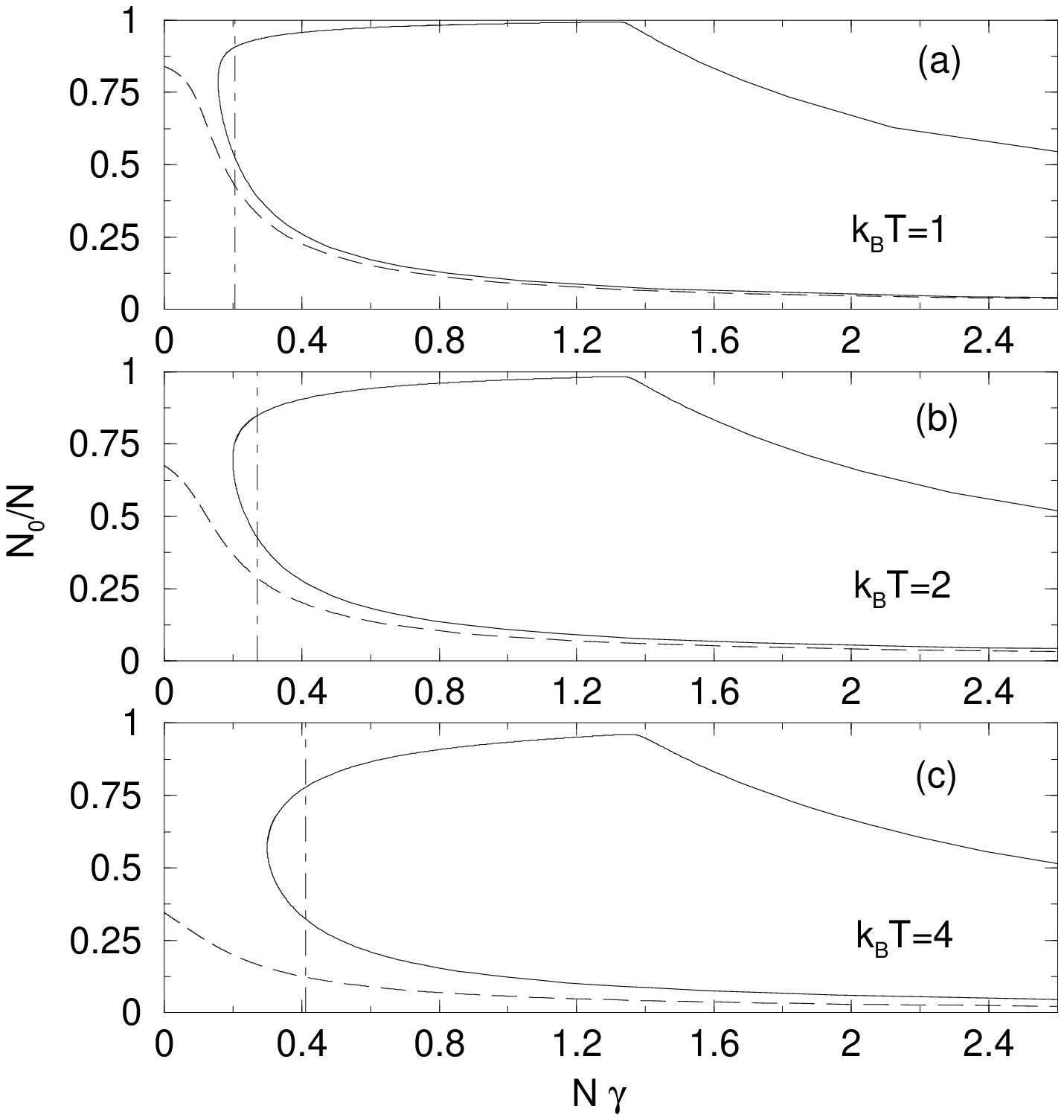,height=3.2in}}
{FIG. 2 Condensate fraction $N_0/N$ as a function 
of the scaled interatomic strength $\gamma = 2|a_s|/a_{\bot}$ 
for $N=10^4$ atoms with $L/a_{\bot}=100$. 
Dashed line: uniform solution. Solid line: 
localized solutions. The vertical dot-dashed line 
indicates the strength $N\gamma_c$ at which 
the uniform-to-localized phase transition takes place.} 
\end{figure}

A coherent bright soliton and a thermal cloud coexist 
in the stable state, as in the metastable localized state. 
This is clearly illustrated 
in Fig. 2 where we plot the condensate fraction $N_0/N$ 
as a function of $\gamma$ for different temperatures. 
At low temperature, see panel (a) of Fig. 2, 
the uniform state (dashed line) is a quasi-pure BEC 
for $0 < \gamma < \gamma_e$. 
For $\gamma >\gamma_e$ the condensate fraction of the uniform 
state quickly decreases. 
At low temperature the uniform-to-localized phase transition 
does not produce a relevant quantum depletion $N_{out}/N$ 
at the transition point. This result is obtained with a fixed 
number of atoms ($N=10^4$). We have verified that 
by fixing $N_0$ instead of $N$ (as done in \cite{jap1}) 
the condensate fraction $N_0/N$ goes to zero for 
$\gamma N_0\simeq \pi^2 a_{\bot}/L$. 
\par 
At higher temperature, 
see panels (b) and (c) of Fig. 2, BEC and non-condensate component 
coexist also in the uniform state. Fig. 2 shows that the condensate 
fraction of the localized states is higher than the condensate 
fraction of the uniform state. Remarkably, at the transition 
strength $\gamma_c$ there is a sizeable jump in the condensate fraction. 
Moreover, for the stable localized state the ratio $N_0/N$ 
initially grows by increasing the interatomic strength. 
As previously discussed, the BEC soliton of the localized state 
will collapse at $\gamma N_0 \simeq 4/3$, but the true collapse 
is avoided by populating the out-of-condensate cloud 
with ${\tilde N}=N-N_0$ atoms for $N\gamma >4/3$. 
We have verified that 
in the stable localized state the number $N_0$ of condensed atoms 
is always very close to the collapse value $4/(3\gamma)$ 
and, for a fixed $\gamma$, the number $N_0$ has a very small 
temperature dependence. It follows that one can estimate 
the condensed fraction of the stable localized state 
as $N_0/N\simeq 4/(3 N\gamma )$. This estimation is not so rough: 
choosing for instance $N\gamma =2$ one finds $N_0/N \simeq 2/3$ 
in good agreement with the result shown in the upper 
solid line of Fig. 2 (see where $N_0/N=0.66$). 
\par 
In the previous calculations, transverse modes have been neglected and
the thermally excited quasi-particles are allowed to populate only the 
lowest transverse state. This assumption can be justified by a simple argument:
In the weakly localized regime (low $N\gamma$) the spectrum of elementary
excitations is well approximated by the free Bose gas, where the logitudinal
modes have energies $O(\hbar\omega_{\bot}a_{\bot}^2/L^2)$ much 
lower than the $O(\hbar\omega_{\bot})$ transverse modes.
In this limit we have explicitly verified that, with the chosen parameters ($L/a_{\bot}=100$), 
for $k_B T \lesssim 5\, \hbar \omega_{\bot}$ the number of quasiparticles 
in the excited transverse modes is indeed negligible. 
In the strongly localized region, both longitudinal and transverse low energy excitations 
are known in the large $L$ limit \cite{jap2,sala02}, where a finite $O(\hbar\omega_{\bot})$ gap
separates the two branches. 
Moreover, our calculations show that for $N \gamma < 4/3$, 
thermal depletion does not affect deeply the thermodynamics of the 
model up to the transition temperature.
\begin{figure}
\centerline{\psfig{file=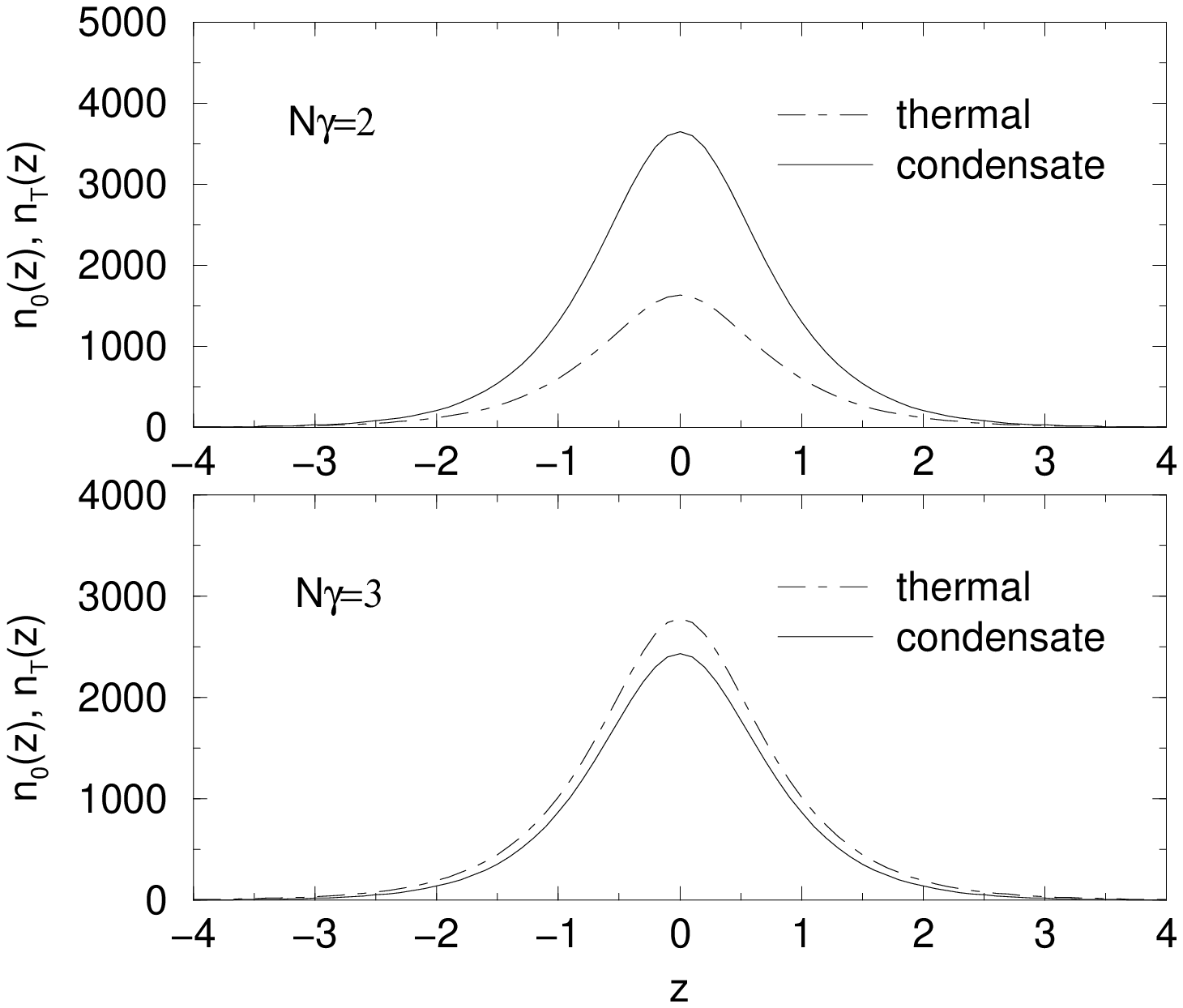,height=2.6in}}
{FIG. 3. Azimuthal density profiles of the 
thermodynamically stable state for $N=10^4$ atoms 
with $L/a_{\bot}=100$ and $k_BT/(\hbar\omega_{\bot})=2$. 
Length $z$ is in units 
of $a_{\bot}=(\hbar/(m\omega_{\bot}))^{1/2}$.} 
\end{figure}
For completeness, we have repeated the calculations by using the 1D GPE. 
Also in this case we have found two localized solutions but 
the stable one always maintains a condensate fraction close to one, 
due to the fact that the 1D GPE does not predict the collapse. 
Thus the transverse structure of the soliton, taken into account 
by the NPSE, strongly modifies the thermodynamics 
of the attractive Bose gas. 
\par 
In Fig. 3 we plot the density profile $n_0(z)$ of the condensate 
cloud (solid line) and of the density profile $n_T(z)$ 
of the thermal cloud (dot-dashed line) for the stable localized state 
and two values of $\gamma$. We have verified that 
the density profile $n_{out}(z)$ of the quantum depletion 
reduces by increasing the temperature $T$ and that 
the quantum depletion $N_{out}/N$ becomes relevant 
only when the $k_B T$ is of the order of the lowest 
Bogoliubov energy level. This energy can be easily 
estimated as $k_B T/(\hbar \omega_{\bot}) =  (\hbar^2/(2m))
(2 \pi/L)^2/(\hbar \omega_{\bot})=  2\pi^2 a_{\bot}^2/L^2 
\simeq 0.002$. 
\begin{figure}
\centerline{\psfig{file=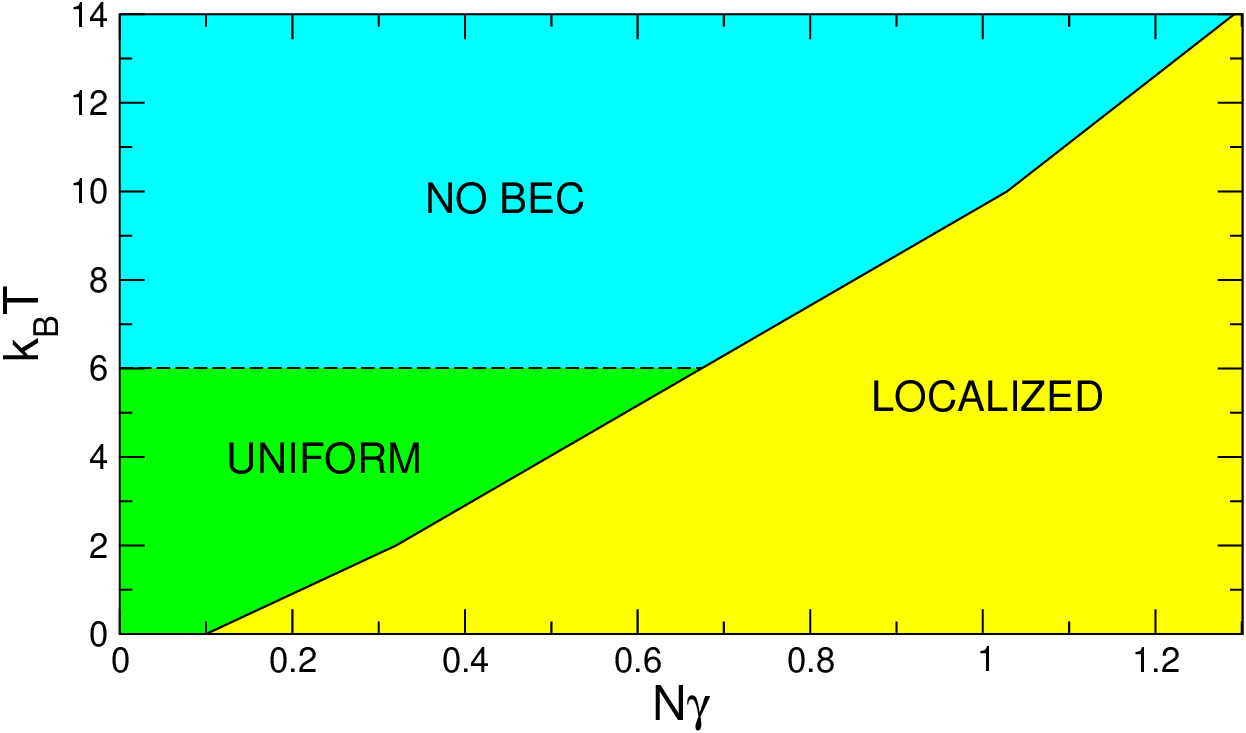,height=1.8in,clip=}}
{FIG. 4 (color online). Phase diagram in the plane 
temperature $T$ vs interatomic strength $\gamma$ 
for $N=10^4$ atoms in a toroidal trap, with $L=100$. 
The energy $k_B T$ is in units of $\hbar \omega_{\bot}$ and 
the length in units of $a_{\bot}$.} 
\end{figure}

\par 
In Fig. 4 we plot the phase diagram of the atomic cloud of bosons 
in the plane $(N\gamma,T)$ for $L/a_{\bot}=100$ and $N=10^4$. 
In the figure we insert a dashed line at 
$k_BT/(\hbar \omega_{\bot}) = 6.01$, which 
is the transition temperature $T_{BEC}$ of the 
Bose-Einstein condensation above which the condensate fraction 
of the uniform state is zero. 
The transition temperature $T_{BEC}$ reduces
by increasing $L/a_{\bot}$. In fact, it is easy to show 
\cite{yukalov} that $k_B T_{BEC}/(\hbar\omega_{\bot})$ 
scales as $N a_{\bot}^2/L^2$. Fig. 4 displays as a solid line 
the curve of the uniform-to-localized transition. 
Remarkably, the BEC transition temperature of the localized solution 
is much larger than the uniform one. 
Fig. 4 shows that with $N\gamma >0.68$, 
there is a first order transition from 
localized BEC phase to the the uniform thermal phase without BEC. 
Instead, for $N\gamma <0.68$ the no-BEC phase is 
obtained starting from the uniform phase by increasing 
the temperature. 
\par 
In conclusion, we have shown that an attractive Bose 
condensate in a quasi one-dimensional toroidal trap 
displays at finite temperature new and intersting features, 
like the coexistence of a coherent self-bound 
bright soliton with a thermal cloud, 
the avoiding of true collapse via population of 
the out-of-condensate (thermal) component, and the enhancement 
of the Bose-Einstein transition temperature for the localized 
solution. We stress that our predictions can be verified 
by using a magnetic ring with geometric parameters 
not too far from those used in Ref.\cite{stamper05} for trapping 
a repulsive condensate of $^{87}$Rb atoms. For instance, 
a sample of $^7$Li atoms, which have a negative scattering 
length ($a_s=-1.4$ nm), can be trapped in a ring with 
transverse width $a_{\bot}\simeq 6$ 
$\mu$m, which gives $\hbar \omega_{\bot}/k_B = \hbar^2/(2 a_{\bot}^2)
\simeq 2$ nK. In this way, having $N=10^4$ atoms the interaction 
strength is $N\gamma =N 2 |a_s|/a_{\bot} \simeq 5$. Then, 
by taking the azimuthal radius $R =L/(2\pi)\simeq 100$ $\mu$m 
one can test the predictions of Fig. 4 varying the temperature 
or tuning the scattering length $a_s$ around a Feshbach resonance.

\end{document}